\documentstyle[12pt, aaspp]{article}
\begin{document}
 
\def\today{\number\year\space \ifcase\month\or  January\or February\or
        March\or April\or May\or June\or July\or August\or
September\or
        October\or November\or December\fi\space \number\day}
\def\fraction#1/#2{\leavevmode\kern.1em
 \raise.5ex\hbox{\the\scriptfont0 #1}\kern-.1em
 /\kern-.15em\lower.25ex\hbox{\the\scriptfont0 #2}}
\def\spose#1{\hbox to 0pt{#1\hss}}
\def\simlt{\mathrel{\spose{\lower 3pt\hbox{$\mathchar''218$}}
     \raise 2.0pt\hbox{$\mathchar''13C$}}}
\def\simgt{\mathrel{\spose{\lower 3pt\hbox{$\mathchar''218$}}
     \raise 2.0pt\hbox{$\mathchar''13E$}}}

\title{Extrasolar Giant
Planets under Strong Stellar Irradiation}
\author{S. Seager \& D. D. Sasselov}
\affil{Astronomy Department, Harvard University, 60 Garden St., Cambridge MA 02138}

\begin{abstract}
We investigate the effects on extrasolar giant
planets [EGPs] of intense irradiation by their parent stars,
describing the issues involved in treating the model atmosphere problem
correctly. 
We treat the radiative transfer in detail, allowing the
flux from the parent star to interact with all relevant depths of the
planetary atmosphere, with no need for a pre-assumed albedo. We
present a low-resolution optical and near-IR
spectrum of a close-in
EGP, focusing on the differences from an isolated planet.

In our dust-free planetary atmospheres we find that Rayleigh
scattering increases the EGP's flux by orders of magnitude shortward
of the \ion{Ca}{2}~H\&K doublet (3930 ${\rm \AA}$), and the spectral
features of the parent star are exactly reflected. In the optical and
near-IR the thermal absorption of the planet takes over, but the
absorption features are changed by the irradiation. The inclusion of
dust increases the reflected flux in the blue; the stellar spectral
lines can be seen blueward of H$\beta$ (4860 ${\rm \AA}$).
\end{abstract}

\section{Introduction}
Within the past 2 years, detection of planets orbiting sun-like stars has 
exploded. To date 9 have been detected, 4 of which orbit amazingly close to 
their parent stars, less than 0.1 AU, which is 4 times closer than
Mercury is to the sun. 
These include 51 Peg b (Mayor \& Queloz 1995), 55 Cnc b, $\tau$ Boo b,
and $\upsilon$ And b (Butler et al. 1997). 
A more distant, yet still close-in EGP, $\rho$ CrB b at 0.23 AU was detected
by Noyes et al. (1997).
Because these planets are being bombarded by intense radiation from their
parent stars, their spectra and temperature structures are
radically altered compared to planets at Jupiter-like distances from their
parent stars.

There is a strong need to theoretically model the close-in EGPs' atmospheres because
ironically they may be the first of the extrasolar planets to be 
detected directly. 
The flux emitted from EGPs represents
only a small fraction of the flux of the parent star at any
wavelength. However, due to dust formed high in their atmospheres, the close-in EGPs' reflected light
in the optical may add significantly
to the observed flux, which may be enough to make the EGPs directly
detectable. For example, Noyes (1998) has developed a spectral separation technique
that uses the Doppler shift of the spectrum of the primary and takes
advantage of the large orbital velocity of the close-in EGPs. 
With reflected light, close-in EGPs may be bright enough for
spectral separation. In addition, models of close-in EGPs are needed for evolutionary and
interior calculations, which are critically sensitive to the detailed
treatment of the planetary
atmosphere.

Recent work on spectral modeling, colors, 
interior modeling, and evolution of EGPs and brown dwarfs (e.g. Burrows et 
al. 1997b; Allard et al. 1997) has been extremely successful.
So far there have been only limited attempts at stellar 
irradiation modeling. Due to the intense radiation field from the nearby star, 
the atmospheres of close-in EGPs must be treated by detailed radiative 
transfer solutions, which is a new problem for planetary physics. In particular, 
planets have outer convection zones, while close-in EGPs develop outer 
radiative zones due to the strong external heating, which inhibits convection 
(Guillot et al. 1996). The first models for irradiated EGPs
were calculated by Saumon et al. (1996) and for close-in EGPs by
Guillot et al. (1996). They assumed for simplicity that the EGPs reflect the 
light of the parent star like a grey body, that the thermal emission of the EGP 
is that of a blackbody, and that a fraction (with a Bond albedo =
0.35) of the flux from the parent star is
absorbed by the planet. Their calculations
also reveal the large sensitivity of the EGPs to the amount of
irradiation.

In this paper we present model atmosphere calculations of the effects
of strong irradiation on an EGP
with particular emphasis on the issues involved in treating irradiation
accurately. As an illustration we 
compute preliminary models for the EGP orbiting $\tau$ Boo.

\section{The Atmosphere Model} 
Our model
atmosphere code is of a type common to irradiation modeling of
close binary stars. The generic version was developed by Nordlund \&
Vaz (1990) for binary stars with $T_{\rm eff}$ of 4500 -- 8000 K, and in
plane-parallel geometry. Our code is made suitable for much cooler temperatures
in the equation of state and opacities. We use
the simplifying assumption of LTE, and the mixing-length theory
for convection. The angle- and frequency- dependent radiative transfer
equation is solved for an
atmosphere with a plane-parallel structure, using the Feautrier method.
Our equation of state includes the following:
up to 2 ionization stages of the elements H,
He, C, N, O, Ne, Na, Mg, Al, Si, S, K, Ca, Cr, Fe, Ni, Ti; the ions
H$^-$, H$_2^+$, H$_2^-$; and the
molecules H$_2$, H$_2$O, OH, CH, CO, CN, C$_2$, N$_2$, O$_2$, NO, NH,
C$_2$H$_2$, HCN, C$_2$H, HS, SiH, C$_3$H, C$_3$, CS, SiC, SiC$_2$, NS,
SiN, SiO, SO, S$_2$, SiS and TiO. The molecular dissociation
constants are from Tsuji (1973).
We include bound-free and free-free atomic opacities from Mathisen
(1984), Thomson scattering, and
Rayleigh scattering by H$_2$ and H. We also include straight means
opacities of H$_2$O (Ludwig 1971) and TiO (Collins \& Fay 1974),
which are the dominant optical and infrared molecular opacities for
$T_{\rm eff}$ of 1500 -- 2000~K.

Our model atmosphere is sufficient for a preliminary 
investigation into the problem of close-in EGPs because we 
are investigating the main effects of stellar irradiation;
we are more
concerned with the correct treatment of radiative transfer of the
incoming radiation than a chemically
complete model atmosphere.

\section{Model Atmospheres for Close-in EGPs}
\subsection{The Radiative Transfer} 
\label{sec-rt}
The radiative transfer problem involves the plane-parallel radiation
field of the parent 
star reaching the upper boundary of the EGP atmosphere and propagating into
it. The atmosphere of the planet can be treated as plane parallel; the
radiative transfer inside it is essentially a 1-D problem with angle
dependence to the normal to the surface. The equation of transfer is
then written as a second-order differential equation with two-point
boundary conditions. It can be solved using Feautrier's method
(e.g. Mihalas 1978), which accounts explicitly for scattering terms
and the two-point boundary conditions --- recovering the diffusion
approximation inner boundary and allowing a simple condition for the
upper boundary. The solution can also treat neighboring frequencies
with very different opacities.
 
Three changes to the Feautrier method are required in order to treat
irradiation of an atmosphere: (1) change of the upper boundary condition to
allow incident radiation from the parent star; (2) in addition
to the angular dependence to the
normal to the surface of the planet, an angular dependence
of the upper boundary condition 
to the azimuthal angle about the normal to the
surface with respect to the direction of the incident flux, ${\em i.e.}$
the incident radiation is not necessarily in the same
plane as the planet's outgoing radiation; and
(3) treatment of incoming radiation by accounting that
the radiation reaches the planet from a specific 
direction, for only some of the same angles as the outgoing
radiation. These last two changes allow for the phases to be correctly
calculated, although for this paper we have assumed that we are
viewing the planet in a
gibbous phase and we have treated the irradiation as flux from the
parent
star, i.e. the star's intensity averaged over angle. This 
small improvement will
be addressed in a later paper.
In general, stellar atmosphere codes that allow
incoming radiation, e.g. MULTI (Carlsson 1992), PANDORA (Avrett \&
Loeser 1992) do not consider
the second two issues,
because they intend to treat incoming radiation from their own
atmosphere, the corona, where the incoming radiation is
only present for the same directions and angles, and in the same plane
as the outgoing radiation. 

The radiative transfer solution with the azimuthal boundary condition
remains essentially the same as the Feautrier method described above, with the
incident and outgoing intensities treated at all appropriate depths.
The radiation temperature of the incoming radiation is not vastly different
(parent stars are solar-like) from that of the close-in EGP, hence very
little frequency reprocessing of radiation occurs.
Solving for the incident specific intensity
involves frequency-dependent processes of absorption, emission,
and scattering, and means no pre-assumed albedo is needed.
Heating is determined by competing processes:
absorption processes (bound-bound, bound-free, free-free)
destroy a
photon, and the photon's energy contributes to the thermal energy of
the gas at the depth at which the photon of that frequency was
absorbed; scattering processes (by
atoms, molecules, dust, etc.) do not contribute to the  kinetic energy
of
particles in the gas and decouple the radiation field from local
conditions.
This detailed radiative transfer approach is equivalent to a
detailed albedo calculation. However, the reverse is not true: model
atmosphere calculations that use albedos (even
frequency-dependent ones), instead of detailed radiative
transfer calculations such as described in this paper, generate
inaccurate
temperature profiles and emergent spectra, because the reflection
and heating from irradiation is a complex function of frequency and
depth. In addition, the reflected light can't easily be separated from
heating or the planet's internal flux. 
For example, the monochromatic
albedo, which we define as the ratio of reflected planetary flux to
incident flux at a specific frequency, is as high as 0.45 around 2400
${\rm \AA}$. In the optical it is less than~0.04, even in our models
that include dust (cf. \S~\ref{sec-emergent}). However, this number
does not illustrate the strong
heating of the planet, which is not included in reflected light.

\begin{figure}
\plotone{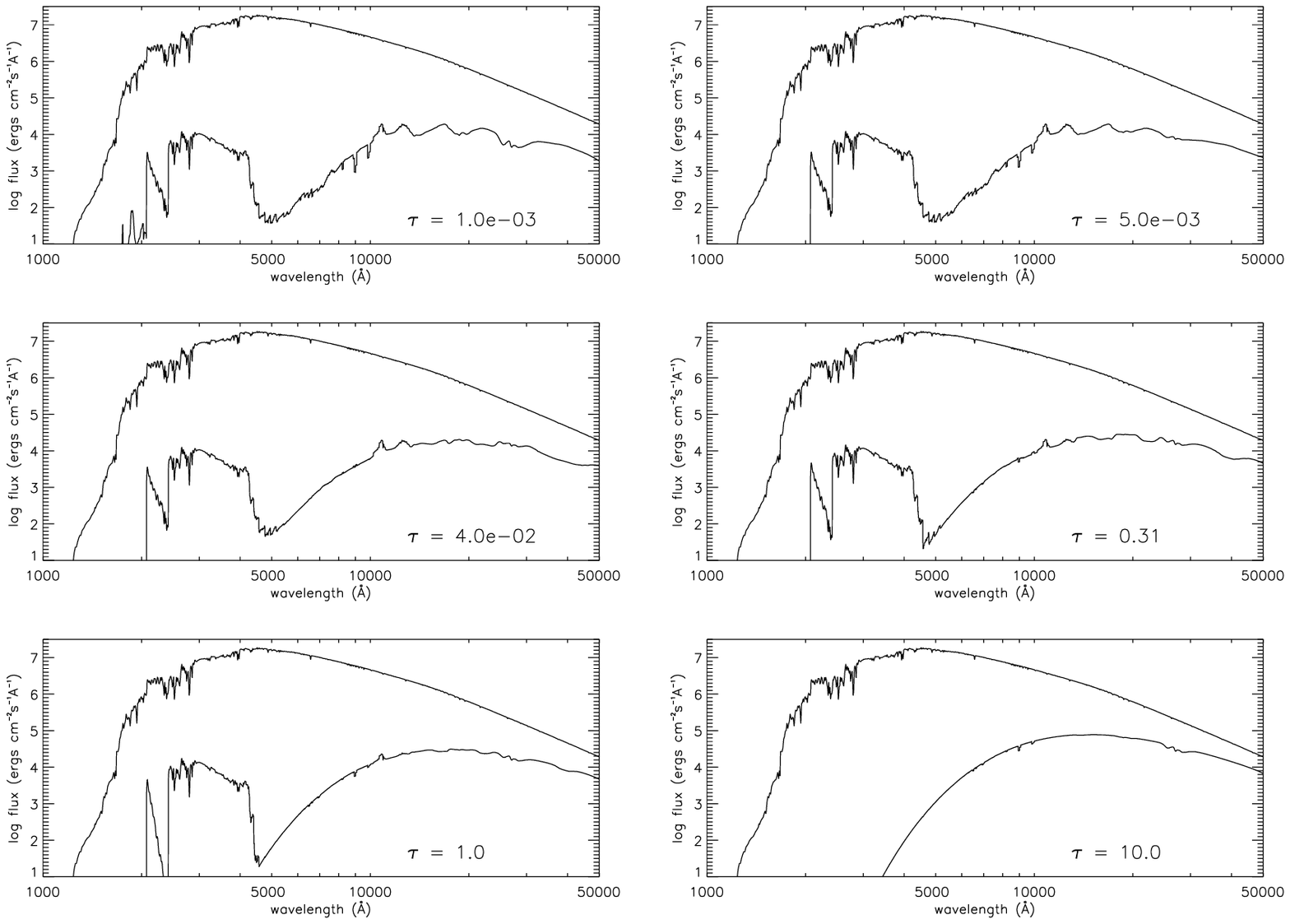}
\caption{Dust-free planetary flux at different optical depths for
$\tau$ Boo
b, for a Rosseland mean optical depth scale. The upper curve in each
frame is the flux of
$\tau$ Boo A. The reflected features
in the near-UV are formed low in the atmosphere, where the density of
H$_2$ is highest.}
\label{fig:fluxwdepth}
\end{figure}

Figure~\ref{fig:fluxwdepth} shows how the radiation penetrates to
different optical depths
depending on frequency. 
The scattered line formation in the
UV range (in the absence of dust) from H$_2$ Rayleigh scattering
occurs very low in
the atmosphere, where the density of H$_2$ is highest. In the TiO
frequency range some of the incoming radiation is absorbed at a much
higher
position in the atmosphere, where the TiO opacity is strongest.
At $T_{\rm eff}$ of 1580 K in this example, in
the absence of dust clouds the incoming radiation in the UV will
penetrate far more deeply than the radiation in the infrared which is
absorbed high in the planetary atmosphere.

\subsection{Conservation of Entropy}
In standard stellar atmospheres flux constancy (of the total radial
energy flux) of all relevant sources is often used to derive the
temperature stratification $T(\tau)$, where $\tau$ is the optical depth.
In a 1-D model calculation of an irradiated planet, we need to abandon total
flux constancy locally in order to conserve entropy at the bottom
of the convection zone. Entropy conservation is required because the
illuminated and non-illuminated atmospheres are two sides of the same planet,
and the interior of both sides will be in the same thermodynamic state.
We assume this is the case for the close-in EGPs, although its
accuracy still needs to be determined. Conservation of entropy at the
bottom of a deep convection zone is a
common approach in irradiation modelling (e.g. Vaz \& Nordlund 1985).
If the irradiation penetrates deeply enough to affect the temperature at
the upper boundary of the convection zone, the effect on the overall
structure of the atmosphere (e.g. $T(\tau)$) is magnified by the
requirement of entropy conservation compared to flux
conservation. This illustrates the need to compute the propagation
of the incident radiation field inside the planet's atmosphere.

\subsection{Temperature Structure Dependence on Radiative
Transfer Methods}
\label{sec-temp}
In the current literature, effective
temperatures ($T_{\rm eff}$) of close-in EGPs are
estimated as the equilibrium effective temperature $T_{\rm eq}$ at a
given distance from
the primary star using an assumed albedo of the planet and the
luminosity of
the primary star (Guillot et al. 1996):
$T_{\rm eq}= T_* (R_*/2D)^{1/2} [f(1-A)]^{1/4}$.
Here the subscript * refers to the parent star, D is the distance of
the planet from the parent star, f=1 if the heat is evenly
distributed, and f=2 if only the heated side reradiates the energy.
Physically, $T_{\rm eq}$ is
the temperature attained by a non-conducting, non-reflecting unit area
perpendicular to the incident flux and located at a distance D from
the parent star.
Because $T_{\rm eq}$ is independent of the radius and mass of the
planet, (dependent only in a complicated manner through the yet
unknown albedo), this definition
reflects the planet's distance from the parent star more than
quantifying the planet's net flux.
During the evolution of an
irradiated planet or for a very massive close-in object, $T_{\rm eff}
> T_{\rm eq}$ due to the
planet's internal flux. While the above equation is a good starting approximation for
$T_{\rm eff}$, more accurate $T_{\rm eff}$ estimates need to be
calculated by
irradiated model atmospheres and interior calculations.

Irradiation can be treated by different methods. Guillot et al. (1996) 
and Saumon et al. (1996) use $T_{\rm eq}$ as the upper boundary condition
of the model atmosphere. A solution would match
this imposed upper boundary temperature in a local (diffusion
approximation) manner. For more details see Guillot \&
Morel 1995. In this way,
all of the radiation is locally coupled to --- and heats --- the
atmosphere, with the heat deposited at the
surface ($T = T_{\rm eq}$) and propagated inward in the gray (constant opacity),
diffusion approximation manner. This method overestimates the change in
the temperature gradient compared to the physical situation because
all incoming radiation is transferred to heat high in the atmosphere.
However, even in the simplest case of an irradiated gray
atmosphere, the temperature stratification is a complex function of
optical depth and incident angle (Vaz \& Nordlund 1985).
In the detailed radiative transfer treatment that we use, some of the
radiation will be scattered
without heating the atmosphere. This is especially important for cooler
close-in EGPs where scattering from dust is efficient. In addition,
the non-blackbody irradiation will penetrate and heat different depths
depending on frequency-dependent absorbers in the planetary
atmosphere. If the radiation propagates fairly deeply
(as shown in Figure~\ref{fig:fluxwdepth}), the entire temperature structure
is significantly changed by the requirement of entropy conservation.

\begin{figure}
\plotone{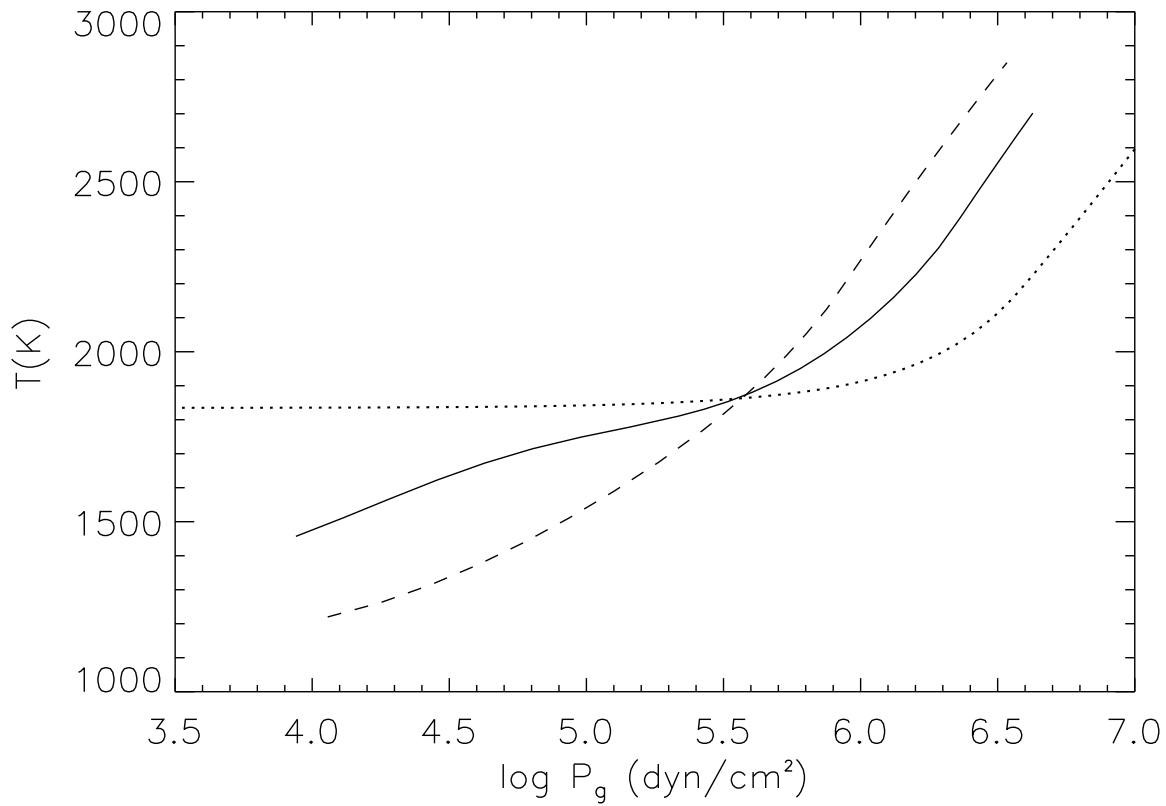}
\caption{Comparison of temperature distributions in the surface layers of
a dust-free model with $T_{\rm eff}$ = 1835 K. 
Two different distributions result
from external irradiation $-$ a fixed external boundary condition model with
$T_{\rm eq}$ (dotted line) and our detailed radiative transfer model (solid
line). They are compared to a brown dwarf atmosphere (same $T_{\rm eff}$)
with no incident external radiation (dashed line). See \S~\ref{sec-temp} for more
details.}
\label{fig:temp1835K}
\end{figure}

\begin{figure}
\plotone{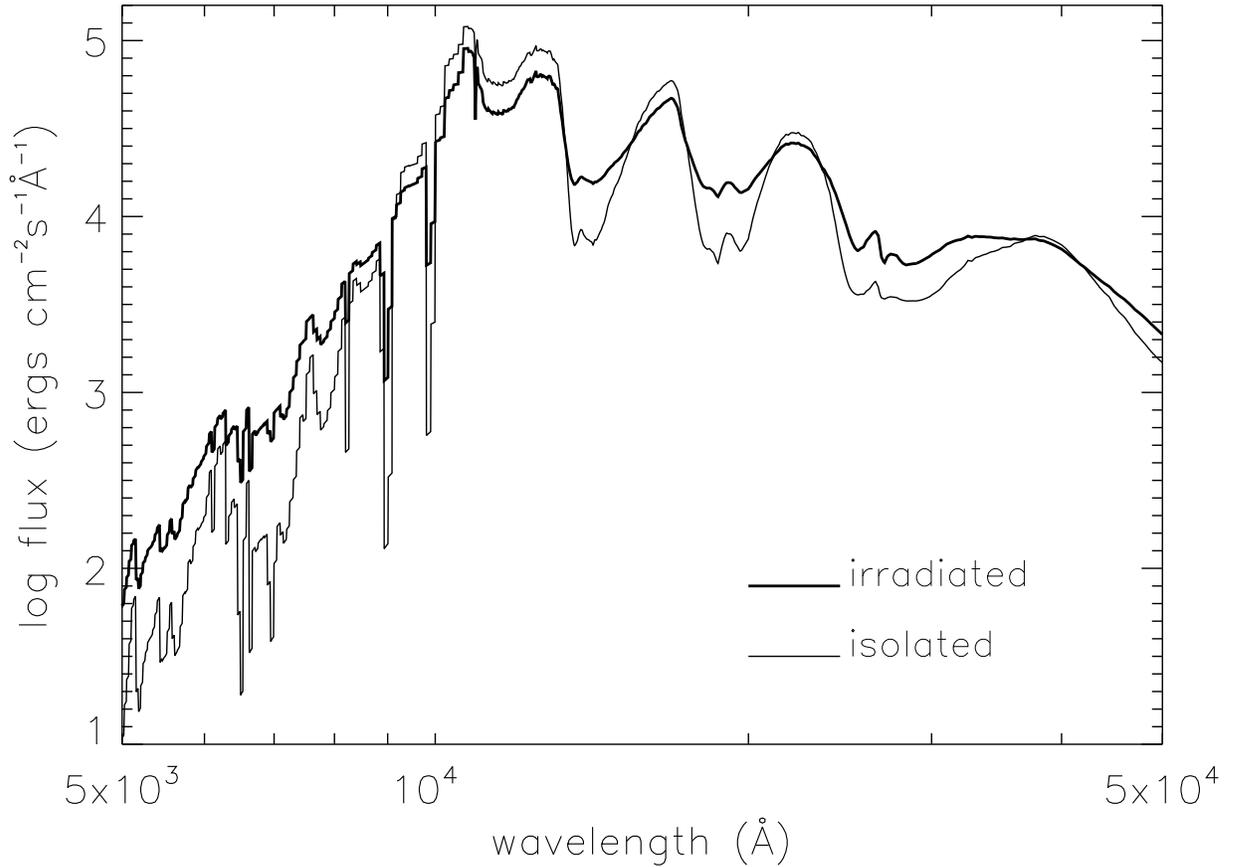}
\caption{Emergent spectra for $T_{\rm eff}$ = 1835 K dust-free 
models of an isolated brown dwarf (thin curve),
and an irradiated planet/brown dwarf (thick curve) at the position of $\tau$
Boo b. The two spectra correspond to the temperature distributions shown in
Figure 2.} 
\label{fig:spectrum1835K}
\end{figure}

The different outcomes are shown for a $T_{\rm eff}=1835$K planet/brown dwarf in
Figure~\ref{fig:temp1835K} which compares the temperature structures of
(a) an isolated planet/brown dwarf,
(b) an irradiated planet (our detailed radiative transfer), and
(c) an irradiated planet with the diffusion approximation method.
Such different temperature distributions
strongly affect the emergent spectral features (as shown in
Figure~\ref{fig:spectrum1835K}).
 
\subsection{Dust}
The strong scattering and absorbing properties of dust will
strongly affect the atmospheres of close-in EGPs, depending
on the location of the dust, usually
resulting in more reflected light in the optical and near-IR.
Ideally, 3-D radiative transfer should be used to compute the dust scattering accurately,
especially since the dust species have very different optical properties
in 3-D whereas in 1-D, the optical profiles of different dust
species may be very similar to each other.
Of further importance to the atmosphere problem,
but not unique to close-in EGPs, dust will ``steal'' away
atoms from the absorbing molecules such as TiO (Tsuji, Ohnaka, \& Aoki 1996),
and in turn reduce their opacities and change the temperature
structure.
 
While our model atmospheres code does not incorporate grain formation,
we illustrate the effect
of silicate dust on the reflection properties of the EGP by
using the
equilibrium condensation range for MgSiO$_3$ described in Lunine
et al. (1989), and scaling its
abundance by that of Mg. We use the Mie theory for opacities
(as described in Ivezic, Nenkova, \& Elitzur 1998) averaged
over all angles, with the interstellar dust grain size distribution as
described in Kim, Martin, \& Hendry (1994) 
(the grain
size distribution in EGPs/BDs is unknown). The optical properties
of the silicates are taken from Ossenkopf, Henning, \& Mathis (1992).

Figure~\ref{fig:spectrum} shows the effect of dust, where it
has reflected spectral features of the parent star in the optical.
For some wavelengths, reflection is superimposed on
absorption features. In
this case the dust reflection dominates the TiO
opacity. With the equilibrium condensation method, very little
silicate dust forms, but still enough to reflect the
incident radiation. While we have taken into account the reduction of
TiO due to O in the silicates, the small amount of dust formation
leaves the TiO and H$_2$O opacities unaffected.
Without dust, scattering in the optical will be minimal,
resulting in little reflected light as shown by the lower
spectrum in Figure~\ref{fig:spectrum} (thick line).

While we do not attempt to address the complex issue of dust
formation, it should be noted that because of irradiation
the close-in EGPs are much hotter
than the observed brown dwarf (BD) Gliese 229B, and Marley et
al. (1998) suggest
there will not be abundant condensates for the temperature range in
which
all close-in EGPs fall.
In general, while model atmospheres that incorporate
dust formation (Allard et al. 1997; Burrows et al. 1997a; Tsuji et
al. 1996) have been very successful so far, many obstacles remain.
Some that are discussed in Allard et al. (1997), Burrows et al. (1997b), and
Lunine et al. (1989) include
accurate dust formation, unknown grain shape and size distribution,
possible clumpy distribution of grains in the atmospheres, and
time-dependent dynamics that remove grains from the upper photosphere over
short time scales.

\section{Emergent Spectrum for a Close-in EGP}
\label{sec-emergent}
\begin{figure}
\plotone{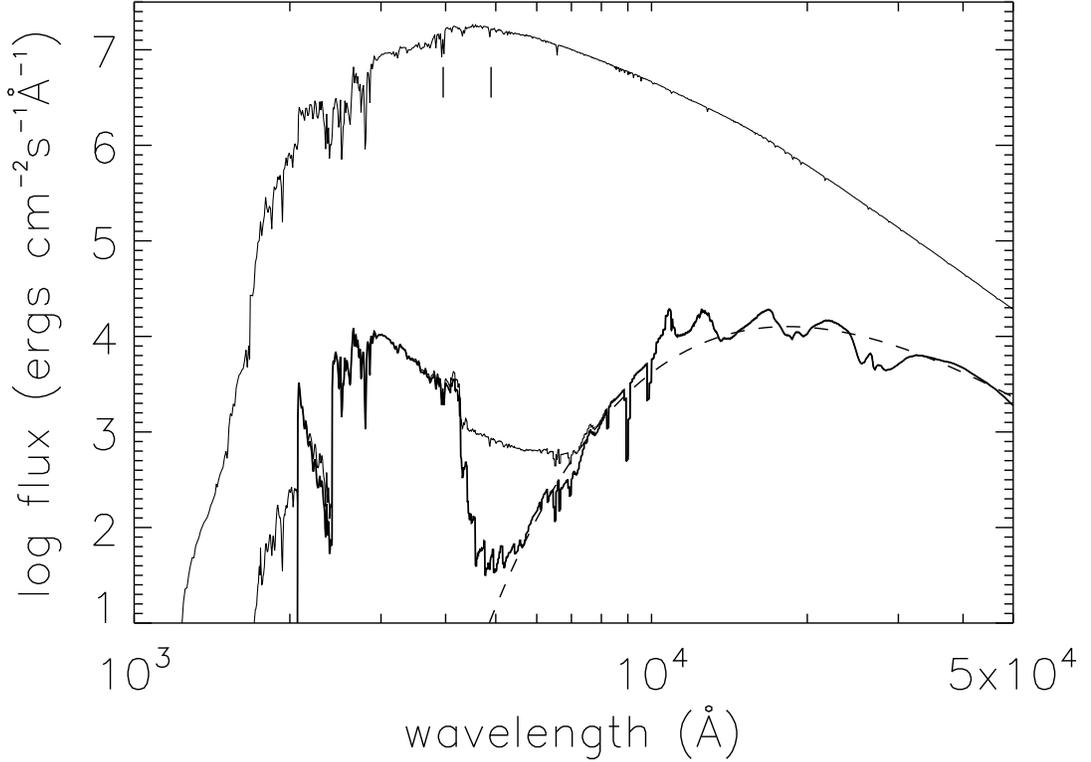}
\caption{Low resolution spectrum (reflected + thermal) of $\tau$ Boo b
(lower thick solid curve),
compared to that of $\tau$ Boo A (upper curve).  The lower thin
curve is $\tau$ Boo b with silicate dust formation, 
where the
reflected spectral lines of the primary can be seen shortward of
H$\beta$ (4860 ${\rm \AA}$). The right tickmark is H$\beta$, the left
is the \ion{Ca}{2}~H\&K doublet; both are seen as reflected features
in the planet. Here $\tau$ Boo b
has $R_B = 1.2 R_J$ and $T_{\rm eff} = 1580 $K; $\tau$ Boo A has a radius
$R_A = 1.2R_{\odot}$. The dashed line is a 1580 K blackbody.}
\label{fig:spectrum}
\end{figure}

As an example of a close-in EGP we use $\tau$ Boo b (Butler et al.
1997), a  planet with mass $M {\rm sin} i$ = 3.87$M_{\rm J}$ and at a
distance
of 0.0462 AU from its parent star. There are many uncertainties about
the close-in EGPs, such as mass, radius, gravity, composition, T$_{\rm
eff }$, etc. We have chosen
only one example out of a large range of parameter space:
$M = 3.87 M_{\rm J}$, T$_{\rm eff }$=1580K, log g (cgs) = 4.0, and
solar metallicity. 
The incident flux of $\tau$ Boo A (F7V, T$_{\rm eff }$=6600, metallicity
[Fe/H] = +0.28, and log g (cgs) =4.5 (Gonzalez 1997)) was taken from
the model grids of Kurucz (1992). 
Figure~\ref{fig:spectrum} shows the emergent spectrum of $\tau$ Boo
b.

We find the following
interesting differences from an isolated EGP:
 
(1) In the spectrum of $\tau$ Boo b (Figure~\ref{fig:spectrum}), the
identical reflected spectral
features from $\tau$ Boo A appear in the near-UV region. In this frequency
range H$_2$ Rayleigh scattering dominates over any absorption
and the result is exact frequency-matched
reflected light. Where
absorption and scattering are comparable, reflected features
from the parent star are seen on top of the 
absorption features from the planet (e.g. from 2065 - 2410 ${\rm \AA}$ in
Figure~\ref{fig:spectrum}).
The abrupt rise in flux below wavelength 4220 ${\rm \AA}$ is likely
due to the wavelength-dependence
of the Rayleigh scattering ($\sim \lambda^{-4}$), although our
opacities are incomplete in that wavelength range.
The existence of silicate dust will result in more
reflected light, as shown in Figure~\ref{fig:spectrum}, especially at
optical wavelengths. 

(2) In wavelength regions of effective scattering the total
(emitted + reflected) flux of $\tau$ Boo b is increased by roughly 1-3
orders of magnitude compared to an isolated planet/brown dwarf (BD) with the 
same $T_{\rm eff}$ of 1580 K. The larger the
temperature difference between the planet and the parent star, the more 
pronounced the increase. 

(3) In an irradiated planet, the TiO and H$_2$O absorption
features are 
shallower and broader due to the reduced temperature gradient in the
upper atmosphere. Figure~\ref{fig:spectrum1835K} shows a comparison of
the emergent spectra for an irradiated and for an isolated planet/BD, both with
$T_{\rm eff}$= 1835 K. 

\section{Conclusion}
This first paper on spectrum formation in close-in EGPs 
shows that strong irradiation 
results in a radically altered temperature
structure and change in absorption features as compared to an isolated EGP with
the same $T_{\rm eff}$. Because spectral features, including
reflection from dust, are highly sensitive to irradiation, model
atmospheres of close-in EGPs must treat the incident radiation
accurately with no need for a pre-assumed albedo. This is especially
relevant for detection
techniques (e.g. Noyes 1998) which use a
narrow range of frequency and detailed information about high
resolution line shapes.


\acknowledgements{
We thank R. Noyes, H. Uitenbroek, and A. Burrows for reading the manuscript and
for useful discussions, and J. Lunine and the referee for valuable comments.
}

\pagebreak

\end{document}